\documentclass[12pt]{article}

\usepackage{amsfonts}
\usepackage{epsfig}

\newcommand{\rf}[1]{(\ref{#1})}
\newcommand{\beq}{\begin{equation}}
\newcommand{\eeq}{\end{equation}}
\newcommand{\bea}{\begin{eqnarray}}
\newcommand{\eea}{\end{eqnarray}}
\newcommand{\e}{\mbox{e}}
\renewcommand{\d}{\mbox{d}}

\renewcommand{\b}{\beta}
\renewcommand{\a}{\alpha}

\newcommand{\eq}{\begin{equation}}
\newcommand{\eqx}{\end{equation}}
\newcommand{\eqn}{\begin{eqnarray}}
\newcommand{\eqnx}{\end{eqnarray}}
\newcommand{\f}[2]{\frac{#1}{#2}}
\newcommand{\eps}{\varepsilon}
\newcommand{\tr}{\mbox{\rm tr}\,}
\newcommand{\Lm}{\Lambda}
\newcommand{\dl}{\delta}
\renewcommand{\th}{\theta}
\newcommand{\al}{\alpha}
\newcommand{\alb}{\bar{\alpha}}
\newcommand{\WW}{{\cal W}}

\newcommand{\nn}{{\cal N}}

\newcommand{\Qt}{\tilde{Q}}
\newcommand{\CC}{{\mathbf C}}
\newcommand{\one}{{\mathbf 1}}

\begin{document}

\begin{center}

\vspace{24pt}

{ \large \bf 

Towards a diagrammatic derivation of the 
Veneziano-Yankielowicz-Taylor superpotential}

\vspace{30pt}

{\sl J. Ambj\o rn}$\,^{a}$
and {\sl R.A. Janik}$\,^{b}$

\vspace{24pt}

{\small

$^a$~The Niels Bohr Institute, Copenhagen University\\
Blegdamsvej 17, DK-2100 Copenhagen \O , Denmark.\\
email: {\tt ambjorn@nbi.dk}\\

\vspace{10pt}

$^c$~Institute of Physics, Jagellonian University,\\
Reymonta 4, PL 30-059 Krakow, Poland.\\
email: {\tt ufrjanik@if.uj.edu.pl}

}

\vspace{48pt}

\end{center}



\begin{center}
{\bf Abstract}
\end{center}

\noindent
We show how it is possible to integrate out chiral 
matter fields in $\nn=1$ supersymmetric theories and 
in this way derive in a simple diagrammatic way
the $N_f S \log S - S \log \det X$ part of the  
Veneziano-Yankielowicz-Taylor superpotential.

\newpage

\subsection*{Introduction}

The recent renewed interest in the calculation of the glueball superpotential
via matrix models \cite{DV} has  led to an understanding  
of how to extract the non-logarithmic part of these superpotentials 
by ordinary diagrammatic methods \cite{DVZ}.
Just as the matrix models in the applications to non-critical strings
and 2d quantum gravity 
were convenient tools for solving  specific combinatorial problems:
the summation over all ``triangulated'' worldsheets with  given weights,
we understand now that the matrix model in the Dijkgraaf-Vafa (DV)
context is an effective way of summing a set of ordinary Feynman
graphs which by the magic of supersymmetry can be combined in 
such a way that they have no space-time dependence.

However, we are still left without a simple diagrammatic derivation
of the logarithmic part of the glueball superpotential, the so called 
Veneziano-Yankielowicz-Taylor superpotential. 
This effective 
Lagrangian was originally derived for a pure $\nn=1$ $U(N_c)$ 
gauge theory by Veneziano and Yankielowicz \cite{VY} by anomaly matching 
and, by the same method, generalized to a $U(N_c)$ theory with 
$N_f$ flavors in the fundamental representation by  
Taylor, Veneziano and Yankielowicz   \cite{VYT}. It is given by 
\eq\label{vyt1}
W_{eff}^{VYT}(S,X) = W_{eff}^{VY}(S)+W_{eff}^{matter}(S,X)
\eqx  
where 
$W_{eff}^{VY}(S)$ is the pure gauge part
\eq\label{e.vy}
W_{eff}^{VY}(S)=-N_c S \log \f{S}{\Lm^3} 
\eqx
while $W_{eff}^{matter}(S,X)$ denotes the part coming 
from $N_f$ flavors in the fundamental representation:   
\eq\label{e.matter}
W_{eff}^{matter}(S,X)= N_f S \log \f{S}{\Lm^3}-S \log\f{\det X}{\Lm^2}.
\eqx
In the above formulas $S$ denotes the composite chiral 
superfield $\WW_\a^2/32\pi^2$ and $X = \Qt Q$
is the ($N_f \times N_f$) mesonic superfield, $Q$ being
the chiral matter field. In \rf{e.vy} and \rf{e.matter} 
$\Lm$ is an UV cut off. Usually this UV cut off is replaced 
by a renormalization group invariant scale $\Lm_M$ by use of the 
one-loop renormalization group:
\beq\label{rgi}
\Lm_M = \Lm \, \e^{-\f{8\pi^2}{(3N_c-N_f)\, g^2}}.
\eeq

The beautiful derivation of \rf{vyt1}-\rf{e.matter} by
anomaly matching has always been somewhat antagonizing since 
a clear diagrammatic understanding is missing. It is summarized
in the following citation from \cite{SV}: ``Its 
[i.e.\ \rf{vyt1}-\rf{rgi}]
only {\it raison d'etre} is the explicit realization of the anomalous
and non-anomalous symmetries of SUSY gluodynamics ....''.

In this letter we point out that there exists a simple diagrammatic 
derivation of \rf{e.matter}. The derivation is inspired by 
diagrammatic techniques used in \cite{DVZ} and the observation that
the DV-matrix models techniques could be extended  
to cover the case of
superpotentials depending on mesonic superfields by considering the
constrained (Wishart) matrix integrals \cite{DJ1}
\eq
\int DQ D\Qt\; \dl(\Qt Q-X)=
\f{(2\pi)^\f{N(N+1)}{2}}{\prod_{j=N-N_f+1}^N (j-1)!} \left( \det X
\right)^{N-N_f} 
\eqx 
and taking the large $N$ limit.

\subsection*{Perturbative considerations}

The matter contribution to the effective superpotential was shown in
\cite{DVZ} to arise from the path integral
\eq
\label{e.pathint}
\int DQ D\Qt \; \e^{ \int d^4 x d^2\th \; 
\left(-\f{1}{2} \Qt( \Box - i \WW^\al
\partial_\al) Q +W_{tree}(\Qt,Q)\right) }
\eqx
where $\WW^\al$ is an external field and $\partial_\al \equiv
\f{\partial}{\partial\th^\al}$. If the quarks are massive
($W_{tree}=m \Qt Q$) then the above path integral reduces to a
functional determinant which can be easily evaluated using the
Schwinger representation:
\eq
\label{e.swdet}
\f{1}{2}\int_{\f{1}{\Lm}}^\infty \f{\d s}{s} \int \f{\d^4p}{(2\pi)^4} \int
\d^2\pi_\al \; \exp \left({-s(p^2+\WW^\al \pi_\al+m)}\right)
\eqx 
where we introduced an UV cut-off $\Lm$. Due to fermionic integrations
the result is
\eq
\label{e.uvcutoff}
\f{\WW^2}{32 \pi^2} \; \int_{\f{1}{\Lm}}^\infty \f{ds}{s} e^{-m s}
\eqx
which reduces for large $\Lm$ to 
\eq
S \log \left(\f{m}{\Lm} \right)
\eqx
At this stage one could integrate-in $X$ to obtain \rf{e.matter}. 
However, as ``integrating-in'' is in fact an 
assumption and we would like to obtain the
desired result perturbatively, or more precisely: diagrammatically. 
To this end we impose the {\em superspace} constraint 
\eq
X=\Qt Q 
\eqx
at the level of the path integral (\ref{e.pathint}). 
This is done by introducing a Lagrange multiplier
chiral superfield $\al$. Since the antichiral sector does not
influence the chiral superpotentials, we will perform a trick analogous
to \cite{DVZ} and introduce an antichiral partner $\alb$ with a tree
level potential $M \alb^2$. Thus we have
\eq
\int \d^4x \d^4\th \;\alb \al +\int \d^4x \d^2\th \; M \alb^2.
\eqx
The path integral w.r.t. $\alb$ is Gaussian and yields
(c.f. \cite{DVZ})
\eq
-\f{1}{2M}\int \d^4x\d^2\th \;{\al \Box \al}.
\eqx
The final path integral is 
\eq
\label{e.pathintfin}
\int D\al D\Qt DQ \; \e^{\int d^4 x d^2\th 
\left(-\f{1}{2} \Qt( \Box - i \WW^\al
\partial_\al) Q - \f{1}{2} \al \Box \al - \al X +\al \Qt Q\right)},
\eqx
where we also took $W_{tree}=0$ and fixed the auxiliary mass $M=1$
(it will be clear from the arguments below that the result is 
independent of $M$).

\begin{figure}
\centerline{\epsfysize=3cm  \epsfbox{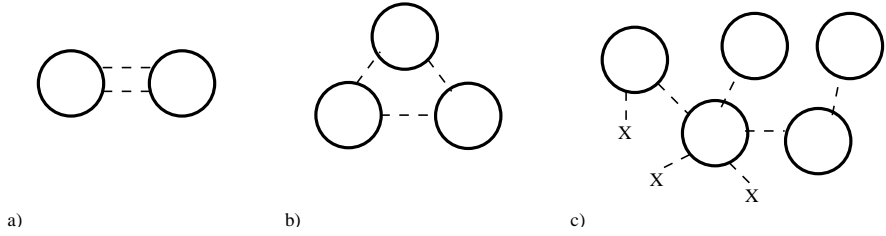}}
\caption{Only tree level graphs survive, i.e.\ we are left with the 
the graphs shown in fig.\ \ref{fig1}c).}
\label{fig1}
\end{figure}

This is no longer a free field theory, but nevertheless there are 
significant simplifications if we only want to extract the 
$\tr \WW^2$ dependence. This implies that we must have 
two $\WW$ insertions per $\Qt Q$ loop. The integrals over the 
fermionic momenta thus force all graphs which
contain an $\al$-line in a loop to vanish. Thus we are left 
with graphs coming from (\ref{e.pathintfin}) which 
have the structure of $\Qt Q$ loops connected 
by at most one $\al$ propagator, and $\al$ propagators
connected to the external field $X$ as shown in fig. \ref{fig1}.

Moreover, if the field $X$ contains a zero momentum component,
which will generically be the case, the integrals will be 
dominated by this constant mode which forces the $\a$ propagators
to be evaluated at zero momentum. 
Consequently we have to introduce an IR cut-off
$\Lm_{IR}$. Each 0-momentum $\al$ propagator will then just contribute a
factor of $1/\Lm_{IR}$.  
Thanks to the above property we may find the full $\Qt Q$ propagator
in terms of the $\al$ 1-point function which we will denote by $F$:
\eq
\f{1}{p^2+\WW^\al \pi_\al+F},
\eqx
and the effective action will be given by the formula (\ref{e.swdet})
with $m$ substituted by $F$:
\eq
\label{e.sprop}
S \log \det \f{F}{\Lm} 
\eqx

\begin{figure}
\centerline{\epsfysize=1cm  \epsfbox{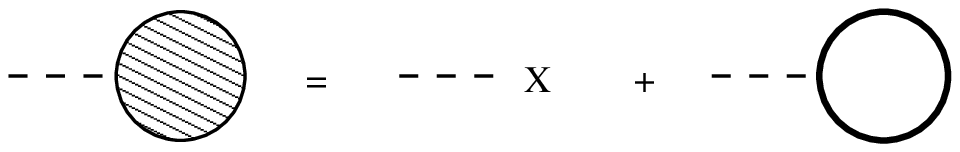}}
\caption{The Schwinger-Dyson equation for $F$.}
\label{fig2}
\end{figure}

It remains to determine $F$. The Schwinger-Dyson equation for $F$ is 
(see fig.~\ref{fig2})
\eq
\label{e.sd}
F=-\f{1}{\Lm_{IR}} X +\f{1}{\Lm_{IR}} \,\f{S}{F}
\eqx
where we used
\eq
\int_{0}^\infty \d s \int \f{\d^4p}{(2\pi)^4} \int
\d^2\pi_\al \;\e^{-s(p^2+\WW^\al \pi_\al+F)} =\f{S}{F}
\eqx
Eq.\ (\ref{e.sd}) is  quadratic and has 2 solutions. Since the
final result has to be IR finite, we will take the solution which has
a finite limit as $\Lm_{IR} \to 0$. Therefore
\eq
F=\f{S}{X}
\eqx
and by substituting this back in (\ref{e.sprop}) one obtains 
the desired result:
\eq
S\log \det\f{S X^{-1}}{\Lm},
\eqx
or, in the case of $N_f$ flavors: 
\beq
N_f S \log \f{S}{\Lm^3} -S
\log \det \f{X}{\Lm^2}.
\eeq

\subsection*{Further examples}

Exactly the same technique can be adapted to the theories studied in
\cite{ILS} where the matter effective superpotentials in terms
of only mesonic fields are quite complex (see eqn. (1.1) in \cite{ILS})
and follow from quite intricate physical analysis. However, as noted
in \cite{ILS} the superpotentials with both  glueball fields 
and matter fields are simpler. The pure matter superpotentials
can then be obtained by integrating out the glueball fields $S_i$.

The simplest case considered in \cite{ILS} is a  
gauge theory with gauge group $SU(2)_1 \times SU(2)_2$,
with a bifundamental matter field $Q$ in the (2,2) representation. 
The natural gauge invariant matter superfield is 
\beq\label{bif}
X = Q^2 \equiv \f{1}{2} Q_{ab} Q_{cd} \eps^{ac} \eps^{bd}, 
\eeq
and the matter part of the superpotential $W_{eff}(S,X)$ is
(eq.\ (4.19) in \cite{ILS}):
\beq\label{weffq}
(S_1+S_2) \log\f{S_1+S_2}{X\Lm} 
\eeq
We will now show that the expression \rf{weffq} also follows from
a diagrammatic reasoning.

Since for $SU(2)$ the fundamental and antifundamental representations
are equivalent through $\Qt_a \equiv Q_{a'} \eps^{a'a}$ the Lagrangian
for the bifundamental fields takes the form:  
\eq
Q_{a' b'}\eps^{a'a}\eps^{b'b}(\Box - i \WW^{(1)\al}_{ac} \partial_\al
- i \WW^{(2)\al}_{bd} \partial_\al )Q_{c d}
\eqx
Again we introduce a Lagrange multiplier superfield $\al$ enforcing the
above constraint. We thus have
\eq
Q (\CC\otimes \CC) (\Box-\WW^{(1)\al} \otimes\one \pi_\al -\one \otimes
\WW^{(2)\al} \pi_\al +\f{1}{2}\al ) Q -\al X
\eqx
where $\CC^{ab} \equiv \eps^{ab}$.

The analogue of formula (\ref{e.sprop}) will then
be
\eq
\label{e.formq}
\f{1}{2} 2(S_1+S_2) \log  \left(\f{F}{2\Lm} \right)
\eqx
where the $1/2$ comes from the fact that we are dealing with a real
representation, while the 2 comes from performing the trace over the
trivial factor in $(\WW^{(1)} \otimes \one)^2$.  
The Schwinger-Dyson equation for $F$ will then have the form
\eq
F=-\f{1}{\Lm_{IR}} X +\f{1}{\Lm_{IR}}\f{1}{2} \f{2(S_1+S_2)}{F/2}
\eqx
hence
\eq
F=\f{2(S_1+S_2)}{X}
\eqx
Inserting $F$ into (\ref{e.formq}) reproduces 
precisely the nontrivial result (\ref{weffq}).

Another example studied in \cite{ILS} for the 
gauge group $SU(2)_1 \times SU(2)_2$ is matter $L{\pm}$ in the 
(1,2) representation. The classical D-flat direction is labeled by 
$Y= L_{\a+}L_{\b-}\eps^{\a\b}$ and the matter 
contribution to  $W^{VYT}_{eff}$ was found in \cite{ILS} to be:
\beq\label{Y}
S_2 \log \f{S_2}{Y \Lm}.
\eeq
We can also reproduce this expression\footnote{Up to a trivial rescaling of
$\Lm$. Note that in our approach the definition of the UV cut-off
$\Lm$ (see e.g. (\ref{e.uvcutoff})) is a matter of convention and may
be modified.} by computing 
diagrammatically the contribution from the $L_\pm$ fields,
starting with the Lagrangian
\eq
L (\CC\otimes \one) (\Box -\WW^{(2)\al} \otimes\one \pi_\al +\a
\one \otimes \CC) L -\a Y,
\eqx
where  the second component in the tensor product is the flavor
space.

\subsection*{Discussion}

We have shown that it is possible to obtain the matter part 
of some generalized $W^{VYT}_{eff} (X,S)$ potentials by simple diagrammatic 
reasoning. It would be interesting to generalize the diagrammatic
derivation to the gauge part of the
Taylor-Veneziano-Yankielowicz superpotential. That would 
complete the diagrammatic derivation of  the glueball
superpotential.

\bigskip

\vspace{12pt}

\noindent{\bf Acknowledgments} 
JA and RJ 
acknowledge support by the
EU network on ``Discrete Random Geometry'', grant HPRN-CT-1999-00161, 
RJ was partially supported by KBN
grant~2P03B09622 (2002-2004) and 
J.A.  partially supported by ``MaPhySto'', 
the Center of Mathematical Physics 
and Stochastics, financed by the 
National Danish Research Foundation.

\end{document}